# A Lattice-Based MIMO Broadcast Precoder for Multi-Stream Transmission

Seijoon Shim, *Member, IEEE,* Chan-Byoung Chae, *Student Member, IEEE,* and Robert W. Heath Jr., *Member, IEEE*

**Abstract**

Precoding with block diagonalization is an attractive scheme for approaching sum capacity in multiuser multiple input multiple output (MIMO) broadcast channels. This method requires either global channel state information at every receiver or an additional training phase, which demands additional system planning. In this paper we propose a lattice based multi-user precoder that uses block diagonalization combined with pre-equalization and perturbation for the multiuser MIMO broadcast channel. An achievable sum rate of the proposed scheme is derived and used to show that the proposed technique approaches the achievable sum rate of block diagonalization with water-filling but does not require the additional information at the receiver. Monte Carlo simulations with equal power allocation show that the proposed method provides better bit error rate and diversity performance than block diagonalization with a zero-forcing receiver. Additionally, the proposed method shows similar performance to the maximum likelihood receiver but with much lower receiver complexity.

**Index Terms**

MIMO systems, broadcast channels, nonlinear system, perturbation methods, spatial filtering, interference suppression.

This work was supported in part by the National Science Foundation under grants CCF-514194 and CNS-435307.

This work was also supported in part by the Korea Research Foundation Grant (KRF-2005-214-D00319) funded by the Korean Government(MOEHRD).

S. Shim, C. B. Chae, and R. W. Heath Jr. are with the Wireless Networking and Communications Group, Department of Electrical and Computer Engineering, The University of Texas at Austin, Austin, TX 78712 USA. (email: {shim, cbchae, rheath }@ece.utexas.edu).

# I. INTRODUCTION

Recent information theoretic work on multiple-input multiple-output (MIMO) communication has shown that the sum capacity, the maximum sum rate in the broadcast channel, is achieved by dirty paper coding (DPC) [1]. The key idea of DPC is to pre-cancel interference at the transmitter using perfect channel state information (CSI) and complete knowledge of the transmitted signals. DPC, while theoretically optimal, is an information theoretic concept that has proven to be difficult to implement in practice. Consequently, several practical near-DPC techniques based on the concept of precoding have been proposed that offer different tradeoffs between complexity and performance [7]–[22].

One of the simplest approaches for multiuser precoding is to premultiply the transmitted signal by a suitably normalized zero-forcing (ZF) or minimum mean squared error (MMSE) inverse of the multiuser matrix channel [7], [8]. The gap in the sum rate between DPC and these linear precoding schemes, however, is quite large due to the transmit power enhancement resulting from power normalization.

A means of avoiding transmit power enhancement is to use non-linear precoding, or lattice precoding [9]–[14], where a modulo operation or vector perturbation is used to reduce transmit power enhancement. The main idea is that an extended constellation is used at the transmitter with multiple equivalent points with the original points in the fundamental constellation boundary. The modulo operation finds a proper point in the fundamental boundary equivalent with a distorted point that the original point moves to in the extended region by power normalization. Tomlinson-Harashima MIMO precoding is one example of transmit precoding with a modulo operation [9], [10]. Another example is vector perturbation where the transmit signal vector is perturbed by another vector to minimize transmit power from the extended constellation [12]. Finding the optimal perturbation involves solving a minimum distance type problem and thus can be implemented using sphere-encoding or other full search based algorithms, which still have moderate complexity. Lower complexity alternatives include lattice-reduction aided broadcast precoding, which uses the Lenstra-Lenstra-Lovasz (LLL) algorithm [13], and a simple vector approximation based on Rayleigh-Ritz theorem [14]. These vector perturbing schemes enable a simple receiver structure via a modulo operation [15]. The multi-user precoding approaches mentioned above assume that the transmitter sends a single stream per each user; in this paper, we consider precoding schemes for multiple stream transmission to increase each user's peak rate in multi-user links.

An alternative to implementing DPC is block diagonalization (BD), which supports multiple stream transmission as well [16]–[19]. The basic concept of BD consists of using special transmit vectors that ensure zero interference between users but do not completely invert the channel. The resulting multiuser MIMO channel matrix has a block diagonal form thus each user can apply a standard point-to-point MIMO receiver. Unlike the aforementioned inverse techniques, BD still requires equalization at the receiver but suffers less from noise enhancement. When user channels are mutually orthogonal, BD achieves the same sum capacity as DPC [20]. The main challenge with BD is that unlike inverse or nonlinear methods, either global CSI is required at all the receivers (obtained through an iterative update for example [18]) or an additional training phase is needed so that each user can estimate their equivalent channel and perform detection [22].

In this paper, we present a lattice-based non-linear precoding scheme that supports multiple stream transmission in a multiuser broadcast channel. All prior approaches mentioned above assume at a minimum complete and perfect CSI at the transmitter. We make the same assumption in this paper. Our proposed scheme exploits the BD linear precoding algorithm to transmit interference free groups of data to different users. To avoid the need for a complex receiver, however, we further use a ZF prefilter combined with a multi-stream vector perturbation to avoid the corresponding power enhancement. The main features of our approach is that (i) we do not require global CSI at the receiver or an additional training phase and (ii) our approach has much lower receiver complexity, at the expense of additional transmit complexity over BD.

We derive the achievable rate of our system under an optimal perturbation assumption. An achievable rate is an error-free supportable rate that satisfies any given power constraint [23]. In our numerical results, we show that the resulting rate is equivalent to that of BD combined with water-filling under an equal power constraint for each user [19], [20]. We also compare the proposed algorithm with previously proposed BD assuming equal power allocation [19] and a ZF or maximum likelihood (ML) receiver [16], [18] in terms of the uncoded bit error rate. We find that our approach has similar diversity performance to BD with an ML receiver and much better performance than BD with a ZF receiver. Thus, from both rate and diversity perspectives, our approach achieves similar performance to BD with an optimal receiver but with much lower receiver complexity. This is a particular advantage in multiuser systems with low-cost low-power mobile users.

This paper is organized as follows. In Section II we begin with the system model and present a summary of BD and its limitations II. In Section III we propose lattice-based precoding with the BD algorithm to support multiple stream transmission and derive its achievable rate. We present numerical results including achievable rate, probability of symbol error, and complexity in IV and conclude in Section V.

## II. Broadcast MIMO System with Block Diagonalization

In this section we discuss the narrow-band broadcast signal and channel model under consideration. Then we discuss block diagonalization and its limitations.

### *A. Notation*

- Let $\mathbf{A}$ denote a complex matrix, and $\mathbf{A}^T$, $\mathbf{A}^H$, and $\mathbf{A}^{-1}$ denote the transpose , conjugate transpose and pseudo-inverse of $\mathbf{A}$, respectively.
- $(\mathbf{a})_l$ and $(\mathbf{A})_{(l,m)}$ denote the $l^{th}$ element of vector $\mathbf{a}$ and the $(l,m)^{th}$ element matrix $\mathbf{A}$, respectively.
- $\text{diag}(a_1, a_2, \cdots, a_n)$ denotes a $n \times n$ diagonal matrix with $\text{diag}(a_1, a_2, \cdots, a_n)_{l,l} = a_l$.
- For a $m \times m$ matrix $\mathbf{A}_l$, $\mathbf{A} = \text{diag}(\mathbf{A}_1, \cdots, \mathbf{A}_n)$ denotes a $mn \times mn$ block diagonal matrix represented by

$$\mathbf{A} = \begin{bmatrix} \mathbf{A}_1 & & \\ & \ddots & \\ & & \mathbf{A}_n \end{bmatrix}.$$

- The trace of a $m \times m$ square matrix $\mathbf{A}$ is expressed as $\text{Tr}(\mathbf{A}) = \sum_{l=1}^{m} \mathbf{A}_{(l,l)}$.

- The Frobenius norm of a $m \times n$ matrix $\mathbf{A}$ is $\|\mathbf{A}\|_F^2 = \text{Tr}(\mathbf{A}\mathbf{A}^H)$.

### *B. MIMO Broadcast Signal Model*

Consider the MIMO broadcast signal model with $K$ users each employing $N_R$ receive antennas and each receiving their own data streams manipulated by a precoder at the base station with $N_T$ antennas as shown in Fig. 1. We assume that the channel is flat fading and for the purpose of simulations we model the elements of each user's channel matrix as independent complex Gaussian random variables with zero mean and unit variance. Such a narrow-band flat fading model is reasonable in future MIMO systems, for example, via orthogonal frequency division multiplexing (OFDM); however, we defer a detailed discussion of OFDM to future work. Let $\mathbf{x}_k$, $\mathbf{H}_k$, and $\mathbf{n}_k$ denote the $k^{th}$ transmit signal vector, the channel from the base station to user $k$, and the thermal noise at user $k$, respectively. The noise $\mathbf{n}_k$ represents additive white Gaussian noise with variance $\sigma_n^2$. In the broadcast channel, since the interference of the other users propagates in the desired user's channel, the received signal at the $k^{th}$ receiver is thus

$$\mathbf{y}_k = \mathbf{H}_k\mathbf{M}_k\mathbf{x}_k + \mathbf{H}_k \sum_{l=1, l\neq k}^{K} \mathbf{M}_l\mathbf{x}_l + \mathbf{n}_k, \tag{1}$$

where $\mathbf{M}_l$ denotes the precoder for the $l^{th}$ user [18]–[22].

### *C. Block Diagonalization and Its Limitations*

In [18] and [19], the authors choose $\mathbf{M}_k$ such that the subspace spanned by its columns lies in the null space of $\mathbf{H}_l$ $(\forall l \neq k)$, that is, $\mathbf{H}_l\mathbf{M}_k = \mathbf{0}$ for $l = 1, \cdots, K-1, K+1, \cdots, K$. If we define $\tilde{\mathbf{H}}_k$ as

$$\tilde{\mathbf{H}}_k = \begin{bmatrix} \mathbf{H}_1^T & \cdots & \mathbf{H}_{k-1}^T & \mathbf{H}_{k+1}^T & \cdots & \mathbf{H}_K^T \end{bmatrix}^T, \tag{2}$$

then $\mathbf{M}_k$ can be obtained by calculating the null space of $\tilde{\mathbf{H}}_k$. Let us define the SVD of $\tilde{\mathbf{H}}_k$ as

$$\tilde{\mathbf{H}}_k = \tilde{\mathbf{U}}_k\tilde{\mathbf{\Lambda}}_k \begin{bmatrix} \tilde{\mathbf{V}}_k^{(1)} & \tilde{\mathbf{V}}_k^{(0)} \end{bmatrix}^H, \tag{3}$$

where $\tilde{\mathbf{U}}_k$ and $\tilde{\mathbf{\Lambda}}_k$ denote the left singular vector matrix and the matrix of ordered singular values of $\tilde{\mathbf{H}}_k$, respectively. Matrices $\tilde{\mathbf{V}}_k^{(1)}$ and $\tilde{\mathbf{V}}_k^{(0)}$ denote the right singular matrices each consisting of the singular vectors corresponding to non-zero singular values and zero singular values, respectively. Note that $\mathbf{H}_l\tilde{\mathbf{V}}_k^{(0)} = \mathbf{0}$ $(\forall l \neq k)$. Assuming each mobile station has $N_{R,k}$ receive antennas, then the $k^{th}$ user can receive $L_k \leq N_{R,k}$ streams. Since the columns of $\tilde{\mathbf{V}}_k^{(0)}$ span the null space of $\tilde{\mathbf{H}}_k$, constructing $\mathbf{M}_k$ using a linear combination of $L_k$ columns of $\tilde{\mathbf{V}}_k^{(0)}$ will automatically satisfy the zero-interference constraints. The specific precoder chosen depends on additional capacity or bit error rate considerations. Assuming that the channel matrices are full rank (which occurs with probability one in complex Gaussian channels), the base station requires that the number of transmit antennas, $N_T$, is at least $\sum_{l=1,l\neq k}^{K} N_{R,l} + L_k$ to ensure there are at least $L_k$ columns in each $\tilde{\mathbf{V}}_k^{(0)}$ and thus satisfy the dimensionality constraint required to cancel interference [19], [21].

When excess transmit antennas are available, i.e., $N_T > \sum_{l=1,l\neq k}^{K} N_{R,l} + L_k$, it is possible to improve BD using transmit antenna selection or eigenmode selection [21]. In addition, when more receive antennas than the number of transmit streams are available, receive antenna selection can further improve BD [20]. In this paper, for notational and analytical simplicity, we assume that every user has the same number of receive antennas $N_R$, the number of transmit data streams makes full use of the receive antennas $L_k = N_R$, and the number of transmit antennas exactly satisfies the dimensionality constraint $N_T = \sum_{k=1}^{K} L_k$.

After pre-canceling the interference of the other users thanks to the precoder $\mathbf{M}_k$, the received signal of the $k^{th}$ receiver, $\mathbf{y}_k$ is given by

$$\mathbf{y}_k = \mathbf{H}_{eff,k}\mathbf{x}_k + \mathbf{n}_k, \tag{4}$$

where $\mathbf{H}_{eff,k} = \mathbf{H}_k\mathbf{M}_k$ denotes the effective channel of the $k^{th}$ user. Since the $k^{th}$ user receives its own data stream without interference from other users, the methodology for designing an appropriate decoder is similar to that for single user MIMO cases after channel estimation [16], [18]. Note that we cannot use a common pilot for estimating $\mathbf{H}_{eff,k}$ since each user uses a different precoding filter $\mathbf{M}_k$ and thus $\mathbf{H}_{eff,k}$ consists of the precoding filter as well as the raw channel $\mathbf{H}_k$ [22]. This means that either an additional training phase or global CSI at the receiver is needed.

To achieve the highest sum rate, after removing the effect of the interfering users' streams, BD

maximizes the data throughput with the well-known water-filling (WF) algorithm [19]. Define the SVD of $\mathbf{H}_{eff,k}$

$$\mathbf{H}_{eff,k} = \mathbf{U}_k \left[\mathbf{\Lambda}_k \ \mathbf{0}\right] \ \left[\mathbf{V}_k^{(1)} \ \mathbf{V}_k^{(0)}\right]^H, \tag{5}$$

where $\mathbf{V}_k^{(1)}$ denotes the set of the right singular vectors corresponding to non-zero singular values and $\mathbf{U}_k$ is the left singular matrix. Assume that $\mathbf{Q}_k$ is a diagonal matrix whose elements scale the power transmitted into each of the column of $\mathbf{M}_k$. When the precoder of the $k^{th}$ user $\mathbf{M}_k$ is given by

$$\mathbf{M}_k = \tilde{\mathbf{V}}_k^{(0)}\mathbf{V}_k^{(1)}\mathbf{Q}_k^{\frac{1}{2}} \tag{6}$$

at the transmitter and the decoder of the $k^{th}$ user also has $\mathbf{U}_k$ at the receiver, the data stream of the $k^{th}$ user is received without the effect of multi-user interference and the maximum achievable sum rate of the BD algorithm, $C_{BD}$, is given by

$$C_{BD} = \max_{\{\mathbf{Q}:\mathrm{Tr}(\mathbf{Q})\leq P_T\}} \log\det\left(\mathbf{I} + \frac{\mathbf{\Lambda}^2\mathbf{Q}}{\sigma_n^2}\right), \tag{7}$$

where $\mathbf{\Lambda}$=diag$(\mathbf{\Lambda}_1, \cdots, \mathbf{\Lambda}_K)$, $\mathbf{Q}$=diag$(\mathbf{Q}_1, \cdots, \mathbf{Q}_K)$, and $\mathbf{Q}(\geq 0)$ denotes the optimal power loading subject to a total power constraint $P_T$ [19]. The optimization problem in (7) is a standard WF problem over the eigenvalues of the equivalent channels [20]. Note that BD achieves the sum capacity achieved by DPC when the user channels are orthogonal (see *Lemma 1*, [20]).

To implement the WF algorithm, knowledge of the decoding filter $\mathbf{U}_k$ is required at each receiver. The decoding filter $\mathbf{U}_k$ though depends on $\mathbf{H}_{eff,k}$, but $\mathbf{H}_{eff,k}$ also consists of the original channel matrix $\mathbf{H}_k$ and the nulling matrix $\tilde{\mathbf{V}}_k^{(0)}$. Because the nulling matrix is calculated by using partial information about the channel state information of other users', the receiver needs to either calculate the decoding filter directly from the estimated channel of $\mathbf{H}_{eff,k}$ [16] or the transmitter can send some information to calculate $\mathbf{U}_k$ at the $k^{th}$ receiver [22].

In the first method, all receivers have to estimate their effective channel including precoding followed by the physical channel. A receiver can estimate the effective channel if dedicated pilot sequences, different for every receiver, are used in the system and precoded by the same transformation. The common pilot, however, is still required, so the first method may increase the control channel overhead. Alternatively, in order to use only common pilot for channel estimation, the transmitter can broadcast the appropriate receiver information to each user [22]. Note that this approach also increase system overhead and new scheme which does not require any coordination information is needed. Therefore, the required coordination is the main limitation of the BD algorithm: all receivers should have global channel state information to generate their own receive filters to approach $C_{BD}$ in (7).

## III. Lattice-Based Broadcast SM-MIMO Precoding System

We introduce a lattice-based (LB) multi-user (MU) spatial multiplexing (SM) MIMO precoding system to support multiple stream transmission in a broadcast channel without any coordination information in this section. Combining BD and perturbation algorithms, we provide a smart solution to avoid coordination information as well as to transmit multiple streams for each user. The proposed scheme requires a simple decoder at each user receiver containing primarily a modulo operation. In this section, we describe the perturbation and our proposed algorithm. Then, we present an achievable rate analysis of the proposed system under the assumption of an optimal perturbation.

### *A. Perturbation Background*

Vector perturbation was introduced in [12] to prevent the transmit power enhancement that occurs when channel inversion using a ZF or MMSE prefilter is used at the transmitter [7]. Prior work considered perturbation in the case where $L_k = 1$, i.e. the single stream case. To help explain our approach we summarize the vector perturbation concept here.

Let $\mathbf{H}$ denote a $K \times N_T$ multi-user channel matrix assuming each user has a single receive antenna. The idea of perturbation is to find a "perturbing" vector $\mathbf{p}$ from an extended constellation ($A\mathcal{C}\mathbb{Z}^K$) to minimize the transmitter power and $\mathbf{p}$ is chosen by solving

$$\mathbf{p} = \arg \min_{\mathbf{p}' \in A\mathcal{C}\mathbb{Z}^K} \left\| \mathbf{H}^{-1} \left( \mathbf{s} + \mathbf{p}' \right) \right\|^2 \tag{8}$$

where $\mathbf{s}$ is a modulated signal vector before perturbing, the scalar $A$ is chosen depending on the original constellation size (we take $A = 2$ for 4-QAM), and $\mathcal{C}\mathbb{Z}^K$ denotes the $K$-dimensional complex lattice[1] [12], [13].

To illustrate, consider the set of equivalent points that form an extended constellation in Fig. 2. A symbol illustrated by the black-filled circle is an original symbol that lies in the fundamental region before pre-distortion by channel matrix inversion ($\mathbf{H}^{-1}$). A set of points marked by the circle is used to represent symbols which are congruent to the symbol in the fundamental region. After pre-distortion, the resulting constellation region also becomes distorted and thus it takes more power to transmit the original point than before distortion because the black-filled point after pre-distortion is further from the origin (+) than before pre-distortion. Note that the set of points marked by the circle represents the same symbol. Among the equivalent points, if the transmitter sends the gray-filled circle point which is the one closest to the origin to minimize transmit power, the receiver finds its equivalent image inside the fundamental constellation region using a modulo operation and treats it as if the black-filled circle point is actually received [15]. Note that the modulo operation is a simply mapping procedure, e.g., any points marked by the circle can be mapped back to the point marked by the black-filled circle via the modulo operation at the receiver in Fig. 2.

The problem of finding the nearest points from the extended constellation is a complex version of the $K$-dimensional integer-lattice least-squares problem [12]. Therefore, an exhaustive search is required to solve (8). It is possible to reduce the search complexity by using lattice reduction algorithms [13], [14].

### *B. Lattice-Based Multiuser Spatial Multiplexing MIMO Precoder Using BD*

The proposed lattice-based multi-user spatial multiplexing MIMO precoding system using the block diagonalization algorithm is illustrated in Fig. 3. The transmitter encodes each user's data streams independently. The $k^{th}$ transmitter consists of the cascade of two filters $\mathbf{H}_{eff,k}^{-1}$ and $\mathbf{M}_k$ where the effective channel $\mathbf{H}_{eff,k} = \mathbf{H}_k\mathbf{M}_k$ calculated as in (4) and the precoding matrix $\mathbf{M}_k$

[1] In [12] and [13], the authors use $2K$-dimensional lattice because they assume that the current realization of $\mathbf{H}$ is separated by real and imaginary values of $\mathbf{H}$. In this paper, however, we use complex version of $\mathbf{H}$ and $K$-dimensional complex lattice defined as $\mathcal{C}\mathbb{Z}^K$.

is given by

$$\mathbf{M}_k = \tilde{\mathbf{V}}_k^{(0)}. \tag{9}$$

Note that $\mathbf{M}_k$ in (9) has a different form from (6). As mentioned in Section II-C, using the optimal BD solution with diagonalization via SVD requires additional coordination information since the effective channel used for the SVD operation includes CSI from other users. To avoid the additional coordination information, the precoder $\mathbf{M}_k$ has only to remove multiuser interference and the inversion of $\mathbf{H}_{eff,k}$ parallelize each user's stream instead of SVD. In addition, we do not require transmit power optimization ($\mathbf{Q}_k^{\frac{1}{2}}$) as seen in (6) since we assume equal power allocation on each stream.

To prevent transmit power enhancement due to $\mathbf{H}_{eff,k}^{-1}$, the proposed transceiver applies a perturbation to the transmitted signal vector to reduce the norm of the precoded signal vector for each user. The perturbation for user $k$ is given by

$$\begin{aligned} \mathbf{p}_k &= \arg \min_{\mathbf{p}'_k \in A\mathcal{CZ}^{L_k}} \left\| \mathbf{H}_{eff,k}^{-1} \left( \mathbf{s}_k + \mathbf{p}'_k \right) \right\|^2 \\ &= \arg \min_{\mathbf{p}'_k \in A\mathcal{CZ}^{L_k}} \left\| \mathbf{H}_{eff,k}^{-1} \tilde{\mathbf{s}}_k \right\|^2 \end{aligned} \tag{10}$$

where $\mathbf{s}_k$ and $\mathbf{p}_k$ are the transmit signal vector of the $k^{th}$ user before perturbing and the perturbing vector of the $k^{th}$ user, respectively. Essentially we find the $k^{th}$ user's perturbing vector $\mathbf{p}_k$ from the set of $L_k$-dimensional complex lattice points. Unlike work in [7], the proposed perturbation operates in stream domain not the user domain. The reason is that the block diagonalization eliminates multiuser interference. In the work in [7], the channel inverse taken on the multiuser channel matrix. In our case, it is taken only over the effective channel matrix after the block diagonalization step.

Since the transmitter sends the pre-distorted symbol with a perturbation, the received signal is given by

$$\mathbf{y}_k = \tilde{\mathbf{s}}_k + \mathbf{n}_k. \tag{11}$$

Note that the received signal in (11) consists of the perturbed symbol ($\tilde{\mathbf{s}}_k$) and AWGN vector. The receiver has only to map the perturbed symbol back to the original symbol ($\mathbf{s}_k$) in the fundamental region using modulo operations [15], and the estimated symbol of $\mathbf{s}_k$ is given by

$$\hat{\mathbf{s}}_k = \mathrm{mod}(\mathbf{y}_k) \tag{12}$$

where $\mathrm{mod}(\cdot)$ denotes a modulo operation. As mentioned in the previous section, $\mathrm{mod}(\cdot)$ results in a simple decoder at the receiver.

*C. Achievable Rate Analysis of MIMO Block Diagonalization with Perturbation*

The problem of achievable rate analysis is reduced to the single-user MIMO case thanks to the fact that the nulling matrix $\mathbf{M}_k$ removes multiuser interference. Recall the received signal model that uses the effective channel in (4). Define $\mathbf{H}_{eff,k} = \mathbf{U}_k \mathbf{\Lambda}_k \mathbf{V}_k^H$ where $\mathbf{U}_k = [\mathbf{u}_1 \ \cdots \ \mathbf{u}_{L_k}]$, $\mathbf{V}_k = [\mathbf{v}_1 \ \cdots \ \mathbf{v}_{L_k}]$ and $\mathbf{\Lambda}_k = \mathrm{diag}\left(\lambda_1 \ \cdots \ \lambda_{L_k}\right)$. Then with $\mathbf{r}_k = \mathbf{U}_k^H \mathbf{y}_k$, $\mathbf{t}_k = \mathbf{V}_k^H \mathbf{x}_k$ and $\mathbf{w}_k = \mathbf{U}_k^H \mathbf{n}_k$, it is possible to transform (4) into the equivalent signal model given by

$$\mathbf{r}_k = \mathbf{\Lambda}_k \mathbf{t}_k + \mathbf{w}_k. \tag{13}$$

Define $\gamma$ as

$$\begin{aligned} \gamma &= \left\| \mathbf{H}_{eff,k}^{-1} \tilde{\mathbf{s}}_k \right\|^2 \\ &= \tilde{\mathbf{s}}_k^H \left( \mathbf{H}_{eff,k} \mathbf{H}_{eff,k}^H \right)^{-1} \tilde{\mathbf{s}}_k \\ &= \tilde{\mathbf{s}}_k^H \mathbf{U}_k \mathbf{\Lambda}_k^{-2} \mathbf{U}_k^H \tilde{\mathbf{s}}_k \\ &= \sum_{l=1}^{L_k} \mu_l^2 \xi_l^2, \end{aligned} \tag{14}$$

where $\mu_l = \frac{1}{\lambda_l}$ and $\xi_l = \left| \mathbf{u}_l^H \tilde{\mathbf{s}}_k \right|$. Since the scalar $\gamma$ is the normalization factor of the transmit signal and $\mathbf{x}_k = \frac{1}{\sqrt{\gamma}} \mathbf{H}_{eff,k}^{-1} \tilde{\mathbf{s}}_k$, the equivalent transmit signal $\mathbf{t}_k$ is given by

$$\mathbf{t}_k = \frac{1}{\sqrt{\gamma}} \mathbf{V}_k^H \mathbf{V}_k \mathbf{\Lambda}_k^{-1} \mathbf{U}_k^H \tilde{\mathbf{s}}_k \tag{15}$$

Substituting (15) for $\mathbf{t}_k$ in (13), the stream-wise form of $\mathbf{r}_k$ is given by

$$\left(\mathbf{r}_k\right)_l = \frac{1}{\sqrt{\gamma}} \mathbf{u}_l^H \tilde{\mathbf{s}}_k + \left(\mathbf{w}_k\right)_l. \tag{16}$$

Now suppose that $\mathrm{E}\{\|\mathbf{x}_k\|^2\} = P$ in (4). From (16), the received signal to noise ratio (SNR) of each stream, $SNR_l$, can be represented by

$$SNR_l = \frac{\rho \xi_l^2}{\gamma} = \frac{\rho \xi_l^2}{\sum_{m=1}^{L_k} \mu_m^2 \xi_m^2}, \tag{17}$$

where $\rho = \frac{P}{\sigma_n^2}$. Therefore, the achievable rate of the $k^{th}$ user, $R_k$, is given by

$$\begin{aligned} R_k &= \sum_{l=1}^{L_k} \log\left(1 + SNR_l\right) \\ &= \sum_{l=1}^{L_k} \log\left(1 + \frac{\rho \xi_l^2}{\sum_{m=1}^{L_k} \mu_m^2 \xi_m^2}\right). \end{aligned} \tag{18}$$

Note that perturbation is not applied yet in calculating (18). Perturbing means that we force the perturbing vector $\mathbf{p}_k$ to minimize $\gamma$ and generate $\tilde{\mathbf{s}}_k$ that can only be coarsely oriented in the coordinate system defined by $\mathbf{u}_1, \cdots, \mathbf{u}_{L_k}$ [12]. Therefore, from (18), if we find the proper perturbing vector and control $\xi_m$ to minimize the normalized factor $\gamma$, then we can obtain the achievable rate of the proposed scheme

$$R_{k,prop} = \sum_{l=1}^{L_k} \log\left(1 + \frac{\rho \xi_l^2}{\min\limits_{\xi_m}\left(\sum_{m=1}^{L_k} \mu_m^2 \xi_m^2\right)}\right). \tag{19}$$

From (19), we need to solve for

$$\xi_m = \arg\min_{\xi_m} \sum_{m=1}^{L_k} \mu_m^2 \xi_m^2. \tag{20}$$

By the Cauchy-Schwartz inequality, solution of (20) occurs when

$$\mu_1^2 \xi_1^2 = \cdots = \mu_{L_k}^2 \xi_{L_k}^2 = \omega_0^2 \tag{21}$$

for an arbitrary constant $\omega_0^2$. Therefore, we calculate the achievable rate $R_{k,prop}$ as

$$R_{k,prop} = \sum_{l=1}^{L_k} \log\left(1 + \frac{\rho \frac{\omega_0^2}{\mu_l^2}}{L_k \omega_0^2}\right) \tag{22}$$

$$= \sum_{l=1}^{L_k} \log\left(1 + \frac{\rho}{\mu_l^2 L_k}\right) \tag{23}$$

$$= \sum_{l=1}^{L_k} \log\left(1 + \frac{\rho \lambda_l^2}{L_k}\right). \tag{24}$$

We obtain (24) substituting $\mu_l$ for $\frac{1}{\lambda_l}$. Note that the solution of (21) is valid when the lattice size is infinite because $\xi_l$ is the relative variable of the perturbed symbol vector $\tilde{\mathbf{s}}_k$ and we would find the proper $\tilde{\mathbf{s}}_k$ provided that the search range of the lattice is infinite. That is, the perturbation finds the perturbed symbol that is the closest point to the origin and also controls the power factor $\xi_l$ to minimize the transmit symbol power.

In addition, (24) is the same expression as the capacity of the single user (SU) MIMO system with equal power allocation, $C_{EQ}$ [24], [25]. Assuming that the elements of each user's channel matrix are identically and independently distributed (i.i.d.), $C_{EQ}$ approaches the capacity of the MIMO system with WF power allocation, $C_{WF}$ for asymptotically high SNR [26]. Thus we conclude that the achievable rate of the $k^{th}$ user, $R_{k,prop}$ obtained by the LB MU SM-MIMO precoding system approaches the optimal capacity that the WF algorithm achieves in a single user MIMO system asymptotically for high SNR, and that the achievable sum rate of the proposed scheme , $R_{sum}$, is defined by

$$\begin{aligned} R_{sum} &= \sum_{k=1}^{K} R_{k,prop} \\ &\Rightarrow \sum_{k=1}^{K} C_{WF,k} \text{ (for high SNR)}, \end{aligned} \tag{25}$$

where $C_{WF,k}$ denotes the achievable rate that each user can approach with the WF algorithm. Note that $\sum_{k=1}^{K} C_{WF,k}$ is a specific solution of (7) with a per-user power constraint that $\text{Tr}(\mathbf{\Sigma}_k) \leq P$ and $KP = P_T$. Consequently, assuming that each user equally uses the transmit power $P$, which is called equal power constraint, this sum rate is the same as the achievable sum rate $C_{BD}$ as shown in (7).

## IV. Numerical Experiments and Results

In this section we compare the sum rate and BER performance of the proposed LB MU SM-MIMO scheme and other various schemes through Monte Carlo simulations. To verify the performance of the proposed LB MU SM-MIMO precoding system, we consider several special cases. For simplicity, without breaking the dimensionality constraints as mentioned in Section II-C, we assume that the number of receive antennas for each user is equal to $N_R$ and that each user receives the same number of streams ($L_k$) as the number of receive antennas, that is, $L_k = N_R$ and that $N_T = KN_R$. We use the notation $\{N_R, K\}$ to index the number of each user's antennas and the number of users. We assume that the elements of each user's channel matrix are independent complex Gaussian random variables with zero mean and unit variance for all numerical results.

### A. Achievable Sum Rate Comparison

We compare the achievable rate of the proposed scheme with the sum capacity, the achievable sum rate with the block diagonalization and the power allocation algorithms and the sum rate of the channel inversion without perturbation.

The sum capacity, $C_{sum}$, denotes the maximum sum rate that can be achieved by DPC [6] given by

$$C_{sum} = \max_{\{\mathbf{R}_k : \sum_{k=1}^{K} \mathrm{Tr}(\mathbf{R}_k) \leq P_T\}} \log\det\left(\mathbf{I} + \frac{1}{\sigma_n^2}\sum_{k=1}^{K}\mathbf{H}_k^H\mathbf{R}_k\mathbf{H}_k\right), \tag{26}$$

where $\mathbf{R}_k(\geq 0)$ is the signal covariance matrix for user k in the dual multiple access channel [3], [6].

We obtain the achievable sum rate of $C_{BD}$ from (7). For the channel inversion scheme, each user exploits the ZF algorithm for precoding after using the nulling matrix and block diagonalization with the constraint that equal power is transmitted to each user's receive antennas without perturbation. To obtain the achievable sum rate of the channel inversion method, the transmit power should be normalized to satisfy the power constraint. Therefore, the achievable sum rate of the channel inversion method, $R_{CI}$, is defined by [19] [27]

$$\begin{aligned} R_{CI} &= \sum_{k=1}^{K} \log\det\left(\mathbf{I} + \left(\frac{\rho}{\left\|\mathbf{H}_{eff,k}^{-1}\right\|_F^2}\mathbf{I}\right)\right) \\ &= \sum_{k=1}^{K} \log\left(1 + \rho\left(\sum_{l=1}^{L_k}\mu_l^2\right)^{-1}\right) \end{aligned} \tag{27}$$

where we recall that $\mu_l$ is the inverse of the $l^{th}$ singular value, $\lambda_l$.

Fig. 4 compares the achievable sum rate of the proposed system with the other systems in the case of $\{2, 2\}$. From Fig. 4, we observe that the sum rate of the proposed scheme is better than that of the channel inversion scheme and also achieves the sum rate of the BD scheme with WF algorithm asymptotically for high SNR as we expected in Section III-C without any additional coordination information and iterative updates for implementing the precoding and decoding filters. The sum rate of the channel inversion scheme is degraded by the power normalization from transmit precoding. The proposed scheme exploits the perturbation as a form of power allocation to compensate for the degradation of power normalization.

Fig. 5 and Fig. 6 show the achievable sum rate performance according to the number of transmit data streams and the number of users, respectively. The sum rate of proposed scheme linearly increases as the number of transmit antennas and the number of users increase, respectively. We observe that performance gap between the proposed scheme and the BD scheme with WF increases as the number of transmit data streams in Fig. 5. The performance gap between $R_{sum}$ and $C_{BD}$ resulted from the assumption that the proposed scheme uses equal power and same constellation for each transmit antenna. Note that the achievable rate of proposed scheme approaches the sum rate of the specific case of $C_{BD}$ mentioned in Section III-C. In Fig. 6, we observe the same tendency that performance gap between the proposed scheme and the BD scheme with the WF increase as the number of users, which is because the number of transmit antennas increases as the number of users increases in this simulation.

### *B. BER performance and Diversity Gain*

In this section we compare the BER performance and diversity gain of the proposed scheme and the other schemes according to several configurations of the receiver antennas and users under assumption that all schemes use equal power and same constellation for each transmit antenna.

The other schemes include ZF MU SM-MIMO, ZF-RX MU MIMO and ML-RX MU MIMO. The ZF MU SM-MIMO scheme removes multi-user interference with the nulling matrix and uses ZF precoding without perturbing the transmit signal as a precoding algorithm. The ZF-RX MU MIMO and the ML-RX MU MIMO schemes are the same as the proposed scheme and the ZF MU SM-MIMO scheme from the viewpoint of the usage of the nulling matrix to remove multi-user interference; however, these schemes have ZF and maximum likelihood (ML) decoders at the receiver to decode the transmit symbol, respectively [16]. Therefore, the receiver uses the coordination information or channel estimation to give the information about the effective channel as mentioned in Section II. We assume that the ZF-RX MU MIMO and the ML-RX MU MIMO schemes exploit perfect channel estimation at the receiver. Note that the precoded channel estimation method is not generally accepted due to the difficulty of designing pilots and preambles in the downlink channel [22]. The proposed scheme and the ZF MU SM-MIMO scheme do not need to estimate precoded channel parameters, which are required in the ZF-RX MU MIMO and ML-RX MU MIMO schemes. In general, all pilot and preambles for channel estimation are not multiplied by any precoder since every user should monitor that pilot and preamble to be served in the near future.

Fig. 7 shows bit error rate (BER) performance comparing the proposed scheme with the ZF MU SM-MIMO scheme for 4-QAM. ZF MU SM-MIMO is the same system as the channel inversion scheme mentioned above. We assume three $\{N_R, K\}$ scenarios to observe the BER performance: $\{2, 2\}$, $\{2, 3\}$, and $\{3, 2\}$. The overall BER performance of the proposed scheme is better than that of the ZF MU SM-MIMO scheme. We have at least 10 dB SNR gain in the proposed system compared with the ZF MU SM-MIMO scheme at $10^{-2}$ BER. From the viewpoint of diversity gain, we also observe that the proposed system has full diversity gain because the proposed system modifies optimal decoders such as the maximum likelihood decoder in transmit precoding and also finds the perturbed symbol which has the optimal decision boundary in the Voronoi region to minimize the transmit power [15]. Therefore, among the BER curves of the proposed LB MU SM-MIMO schemes, the case with $N_R = 3$ shows better diversity gain than the one with $N_R = 2$.

Fig. 8 also shows BER performance comparing the proposed scheme with the ZF-RX MU MIMO and ML-RX MU MIMO schemes for 4-QAM. The proposed scheme supports 9 dB SNR gains at $10^{-2}$ BER compared with the ZF-RX MU MIMO scheme. Compared with the ML-RX MU MIMO scheme, the proposed scheme shows the same diversity gain, but provides less performance in SNR gain. Note that the ML-RX MU MIMO scheme requires perfect channel estimation for decoding the transmit symbol. On one hand, as long as perfect channel estimation is guaranteed at the receiver, ML decoding is the optimal solution to minimize BER. On the other hand, the proposed scheme shows comparable BER performance to the ML-RX MU MIMO scheme without any channel estimation. From the viewpoint of diversity gain, the proposed scheme has the same diversity order with the ML type receiver as mentioned earlier.

We observe the performance in terms of diversity gain particularly in Fig. 9. The simulation results shows the BER performance in the case of $\{3, 2\}$. We observe that the proposed scheme has the same slope as the ML-RX MU MIMO scheme with the ML receiver and provides better diversity gain than the ZF MU SM-MIMO and ZF-RX MU MIMO schemes with linear precoding and decoding, respectively.

### *C. Complexity*

In this section we calculate approximate complexities of all schemes mentioned in Section IV-B.

It is hard to calculate exact complexity of the proposed scheme because the perturbing algorithm used in the proposed scheme adopts a scalar design parameter $\tau$ that provides a symmetric encoding region around every signal constellation points (see (9) in [12]). Since the proposed perturbing algorithm uses a complex version of $L_k$-dimensional integer-lattice least-square problem and we assume that $N = L_k = N_R$, the proposed scheme has an approximate complexity of $\mathcal{O}(N^{M+\alpha})$ for each user, where $M$ denotes a modulation order and $\alpha$ is a positive value that depends on the encoding region parameter $\tau$. The complexity of $\mathcal{O}(N^{M+\alpha})$ is greater than $\mathcal{O}(N^{M})$ referred to as the complexity of ML decoding scheme with $L_k$ transmit data streams. Consequently, the proposed scheme has the complexity of $\mathcal{O}(KN^{M+\alpha})$ ($> \mathcal{O}(KN^{M})$) totally at the transmitter. The receive complexity of the proposed scheme depends primarily on the modulo algorithm that simply demaps a perturbed symbol to an original symbol. Therefore, the proposed scheme has a complexity of $\mathcal{O}(N)$ for each user's receiver.

The ZF-RX MU MIMO and the ML-RX MU MIMO schemes has no special encoding techniques at the transmitter, however, they use the ZF and the ML decoding algorithms at the receiver, respectively. Therefore the approximate complexity orders of ZF-RX MU MIMO and ML-RX MU MIMO schemes are $\mathcal{O}(N^{\omega})$ ($2 < \omega < 3$, see [21]) and $\mathcal{O}(N^{M})$ for each user's receiver, respectively.

We summarize the characteristics of the proposed scheme, ZF MU SM-MIMO, ZF-RX MU MIMO and ML-RX MU MIMO from the viewpoints of channel estimation and transmitter-receiver complexity in Table I.

## V. Conclusions

We have proposed a lattice-based broadcast spatial multiplexing MIMO precoding scheme that supports multi-stream transmission without any coordination information. The proposed lattice-based multi-user precoder uses the block diagonalization scheme to remove multi-user interference and the channel inversion algorithm applied for the calculated effective channel as the precoding algorithm to avoid additional coordination information. It also exploits a perturbation to reduce transmit power enhancement resulting from the power normalization when channel inversion algorithms such as zero-forcing and minimum mean squared error algorithms are used for the precoding algorithm. The proposed scheme can achieve a sum rate approached by the block diagonalization scheme with the water-filling algorithm asymptotically for high signal to noise ratio without any coordination information. Through Monte Carlo simulations, we verified that the proposed scheme has at least 10 dB SNR gain compared with the zero-forcing multi-user MIMO precoding system with the block diagonalization at $10^{-2}$ BER. Also, the proposed scheme provided 9 dB SNR gain at $10^{-2}$ BER compared with the ZF-RX MU MIMO scheme which assumes perfect channel estimation and uses the zero-forcing algorithm at the receiver. Furthermore, we observed that the proposed system gets the full diversity gain, which also holds for the optimum decoding system and achieves the sum rate that the block diagonal scheme with water-filling algorithm asymptotically assuming that the elements of each user's channel are identically and independently distributed. We leave the reduced-complexity implementation of the proposed precoding scheme for future work.

TABLE I

THE SYSTEM CHARACTERISTIC COMPARISON BETWEEN THE PROPOSED SYSTEM, ZF MU SM-MIMO, ZF-RX MU MIMO AND ML-RX MU MIMO

| | Channel estimation * | Transmitter complexity | Receiver complexity ** |
|---|---|---|---|
| The Proposed Scheme | No | $> \mathcal{O}(KN^M)$ | $\mathcal{O}(N)$ |
| ZF MU SM-MIMO | No | $\mathcal{O}(KN^\omega), 2 < \omega < 3$ | $\mathcal{O}(N)$ |
| ZF-RX MU MIMO | Required | - | $\mathcal{O}(N^\omega), 2 < \omega < 3$ |
| ML-RX MU MIMO | Required | - | $\mathcal{O}(N^M)$ |

* This channel estimation means estimation for the effective channel at each mobile station.

** This receiver complexity is required for each mobile station.

*** We assume that $N = N_R = L_k$, and $M$ denotes a modulation order.

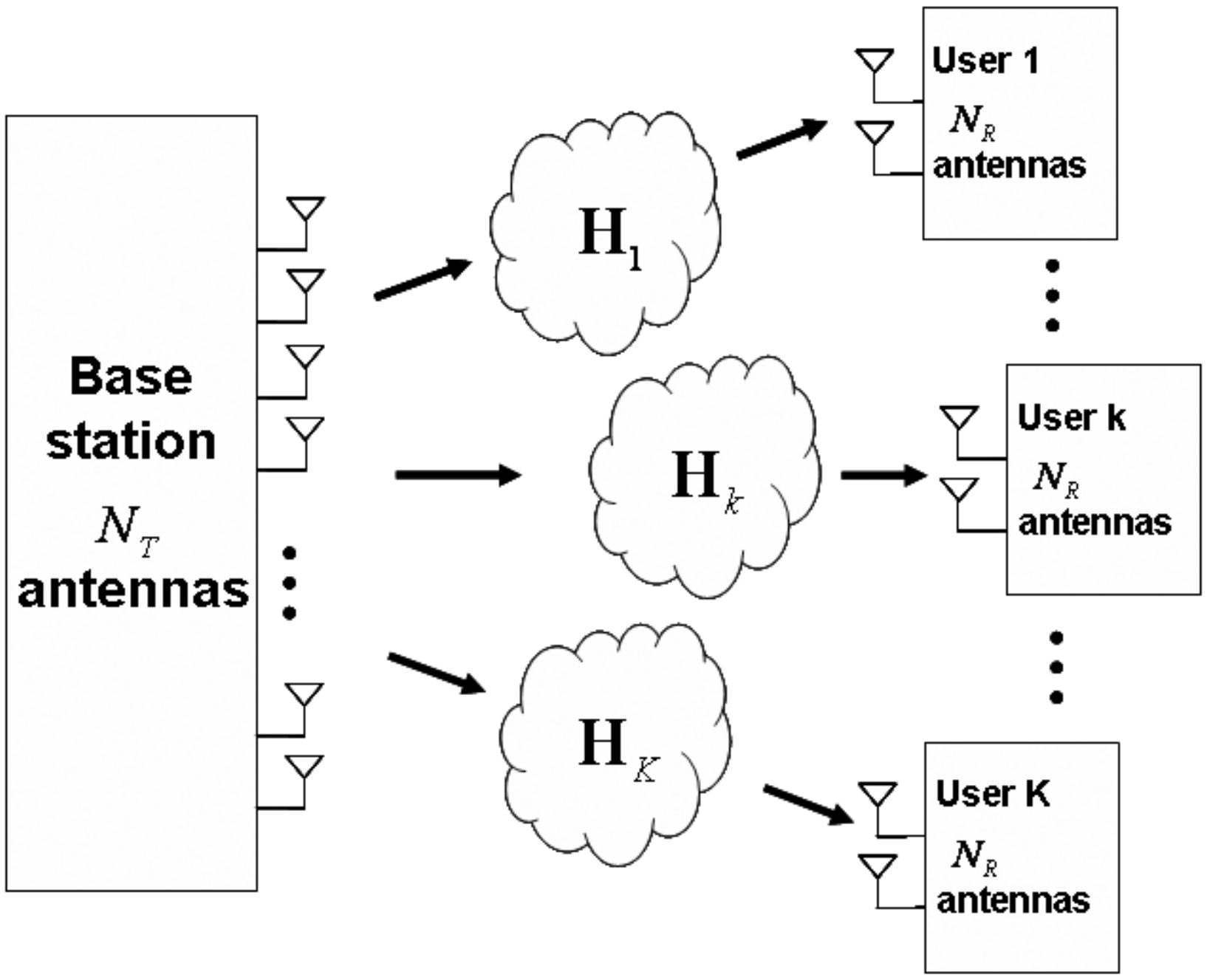

Fig. 1. A MIMO broadcast system where a base station and each user has $N_T$ transmit antennas and $N_R$ receive antennas, respectively.

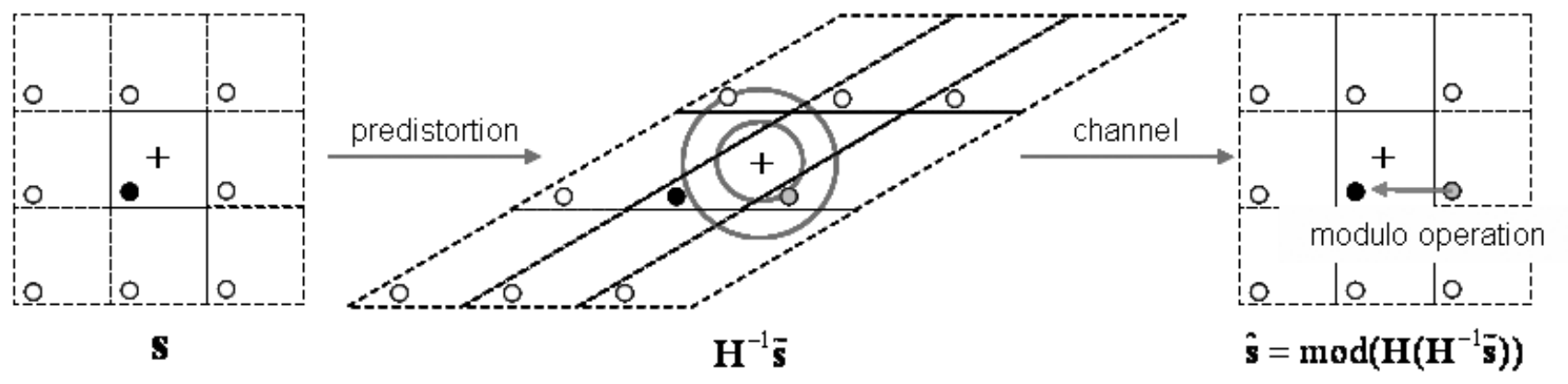

Fig. 2. The concept of perturbation.

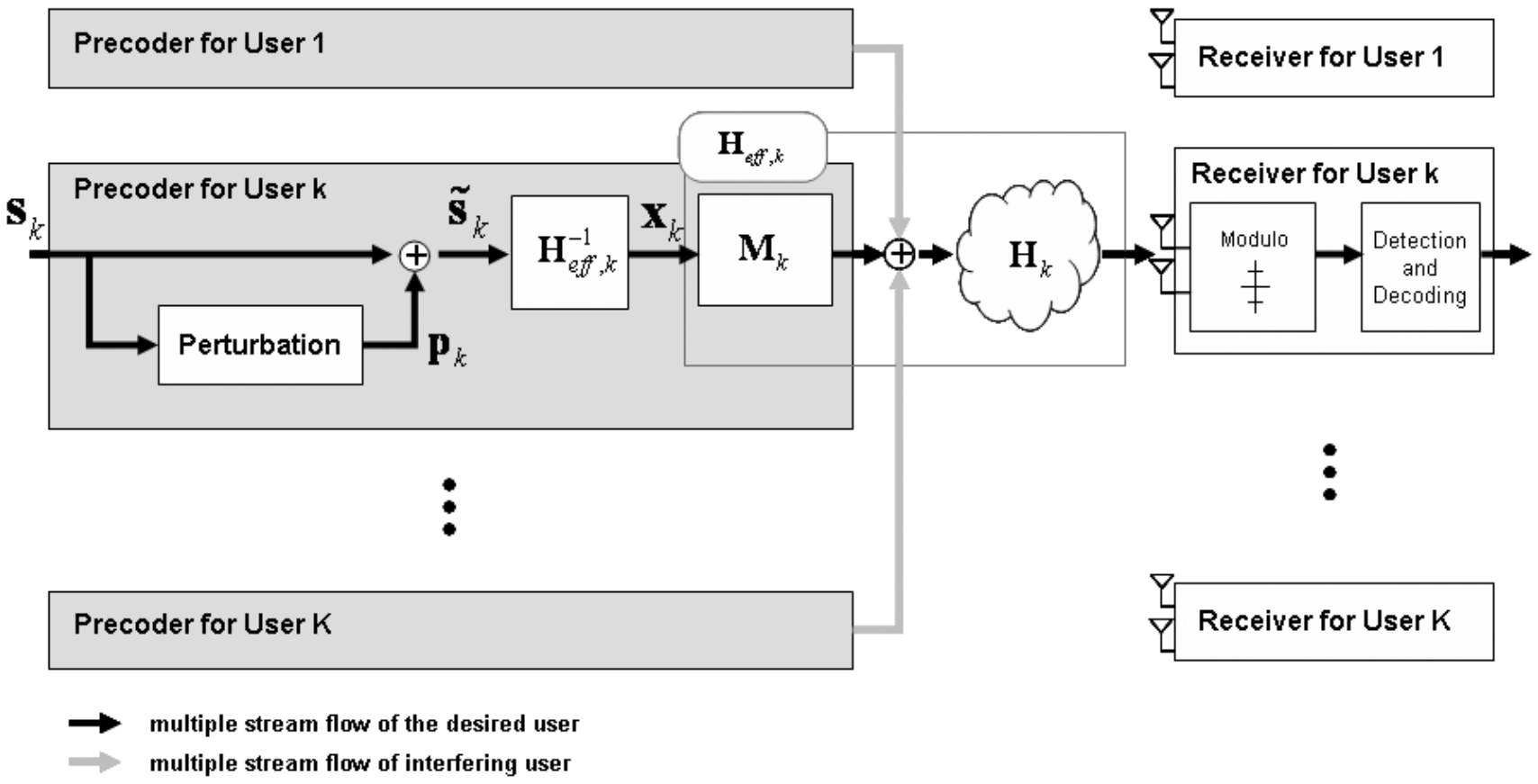

Fig. 3. The structure of a lattice-based broadcast SM-MIMO precoding system using the block diagonalization algorithm.

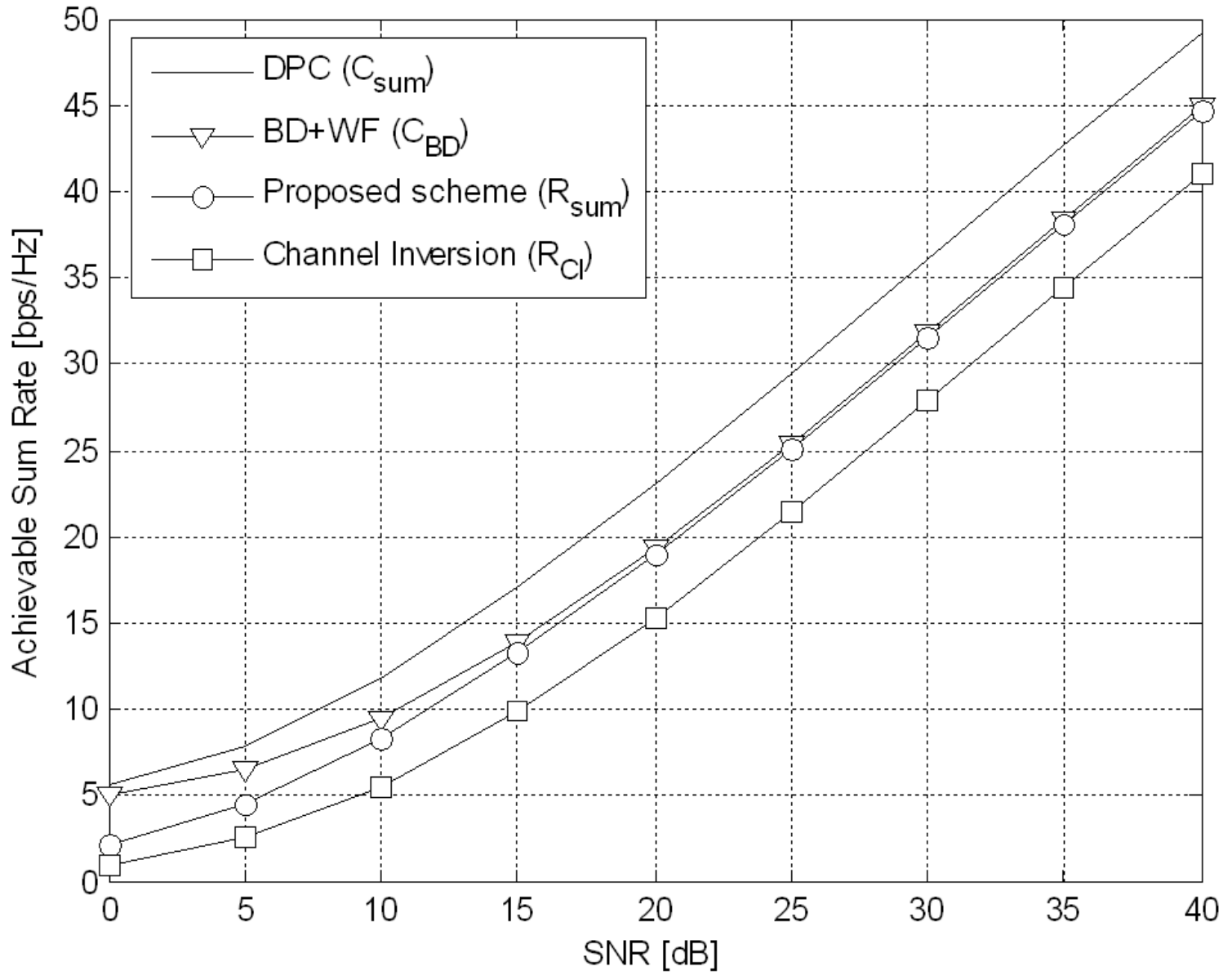

Fig. 4. The comparison between the sum-capacity ($C_{sum}$, [8]) and the achievable sum rates of the BD scheme with WF ($C_{BD}$, [19]), the channel inversion scheme ($R_{CI}$) and the proposed scheme ($R_{sum}$). $N_T = 4$, $N_R = 2$ and $K = 2$.

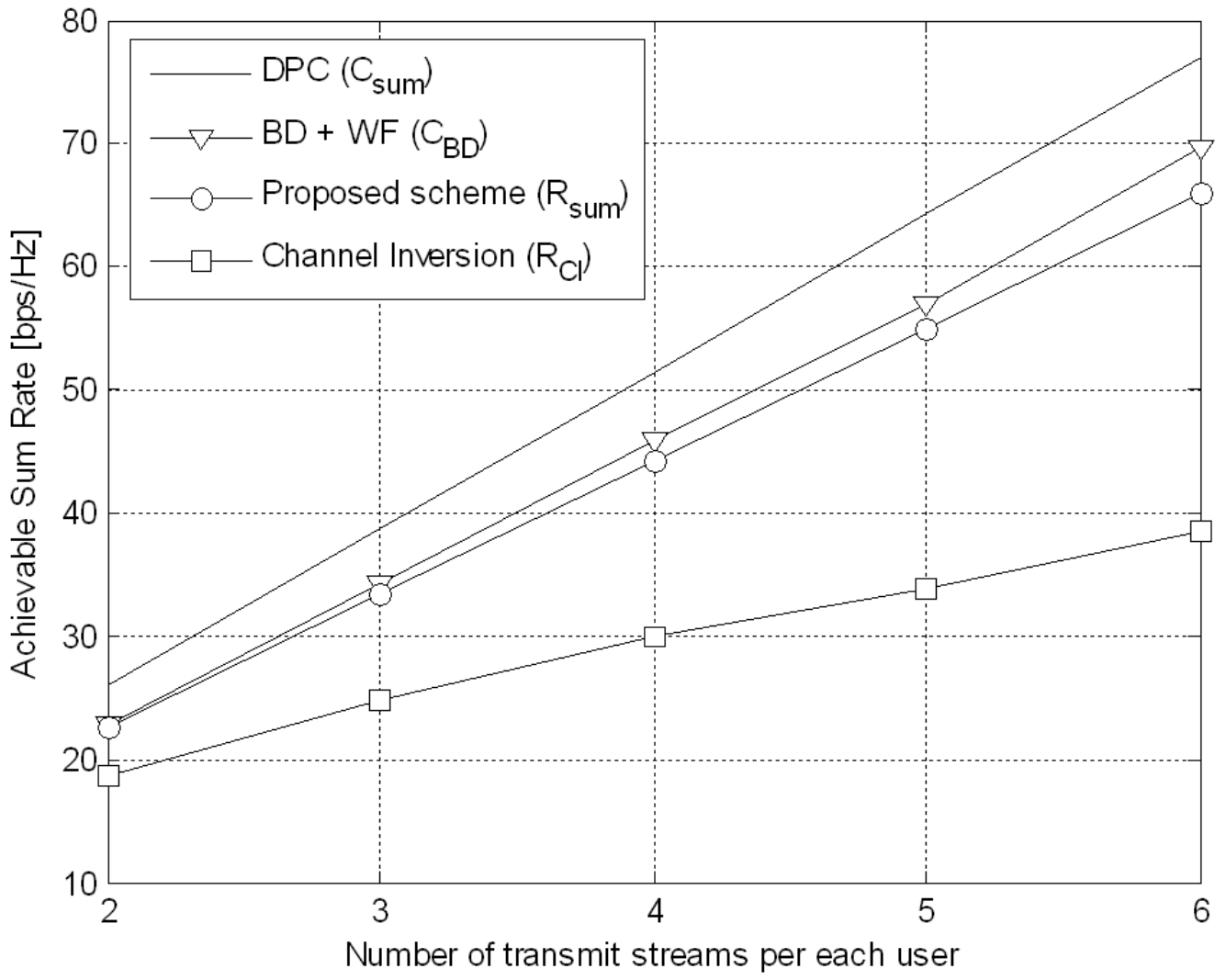

(a)

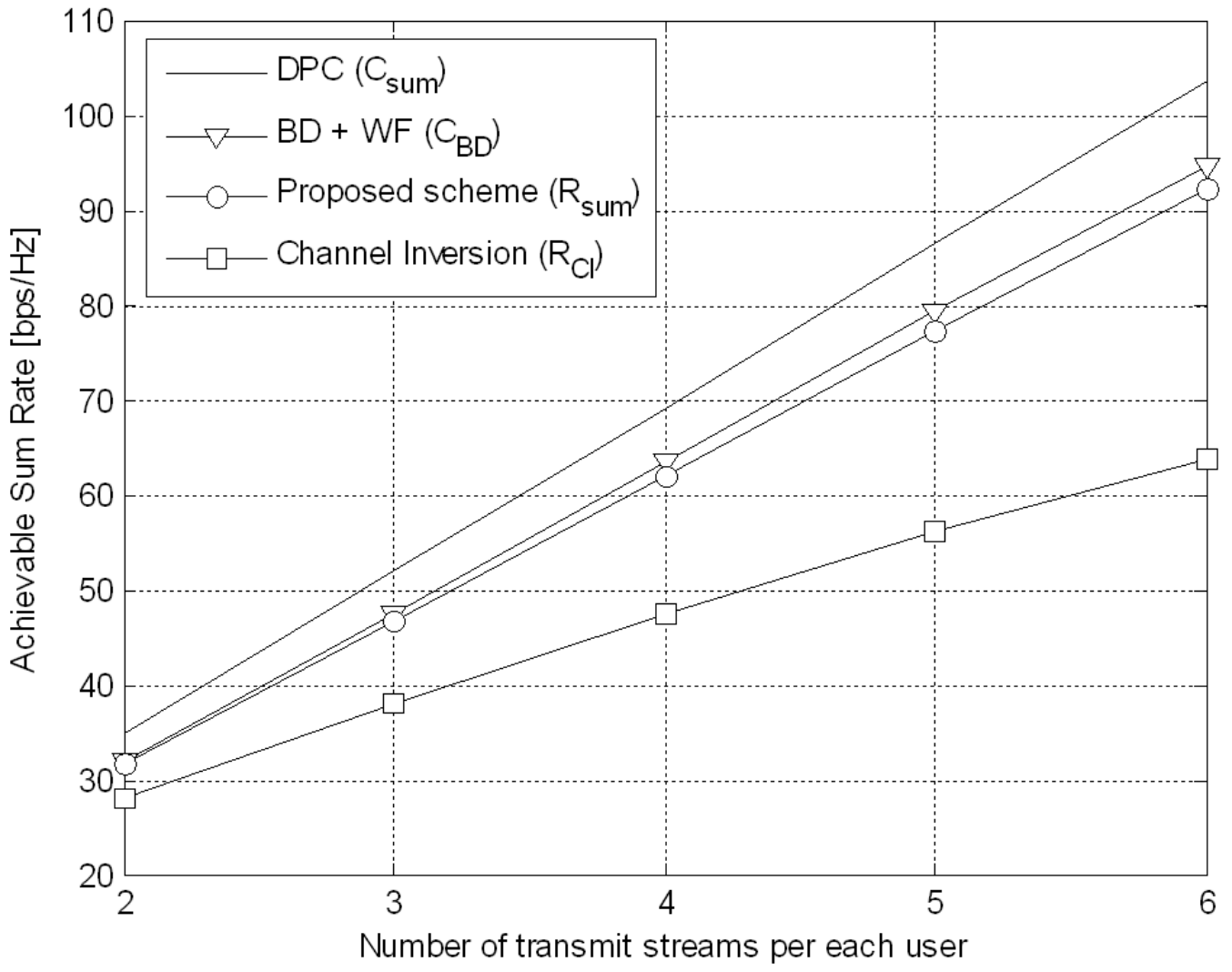

(b)

Fig. 5. The achievable sum rate performance according to the number of transmit streams per each user: K=2, (a) SNR=20dB, (b) SNR=30dB.

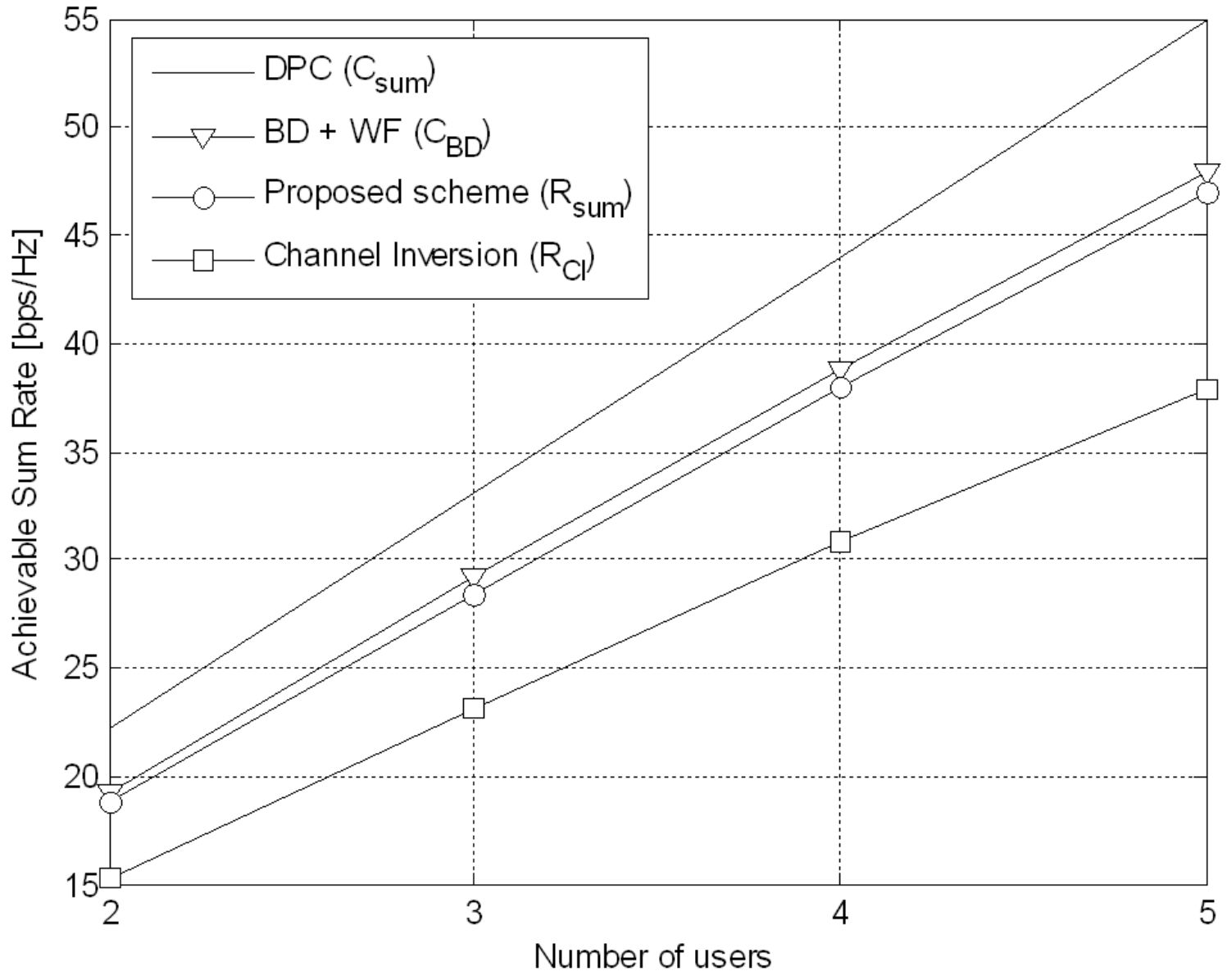

Fig. 6. The achievable sum rate performance according to the number of users: $N_R = 2$ and SNR=20dB.

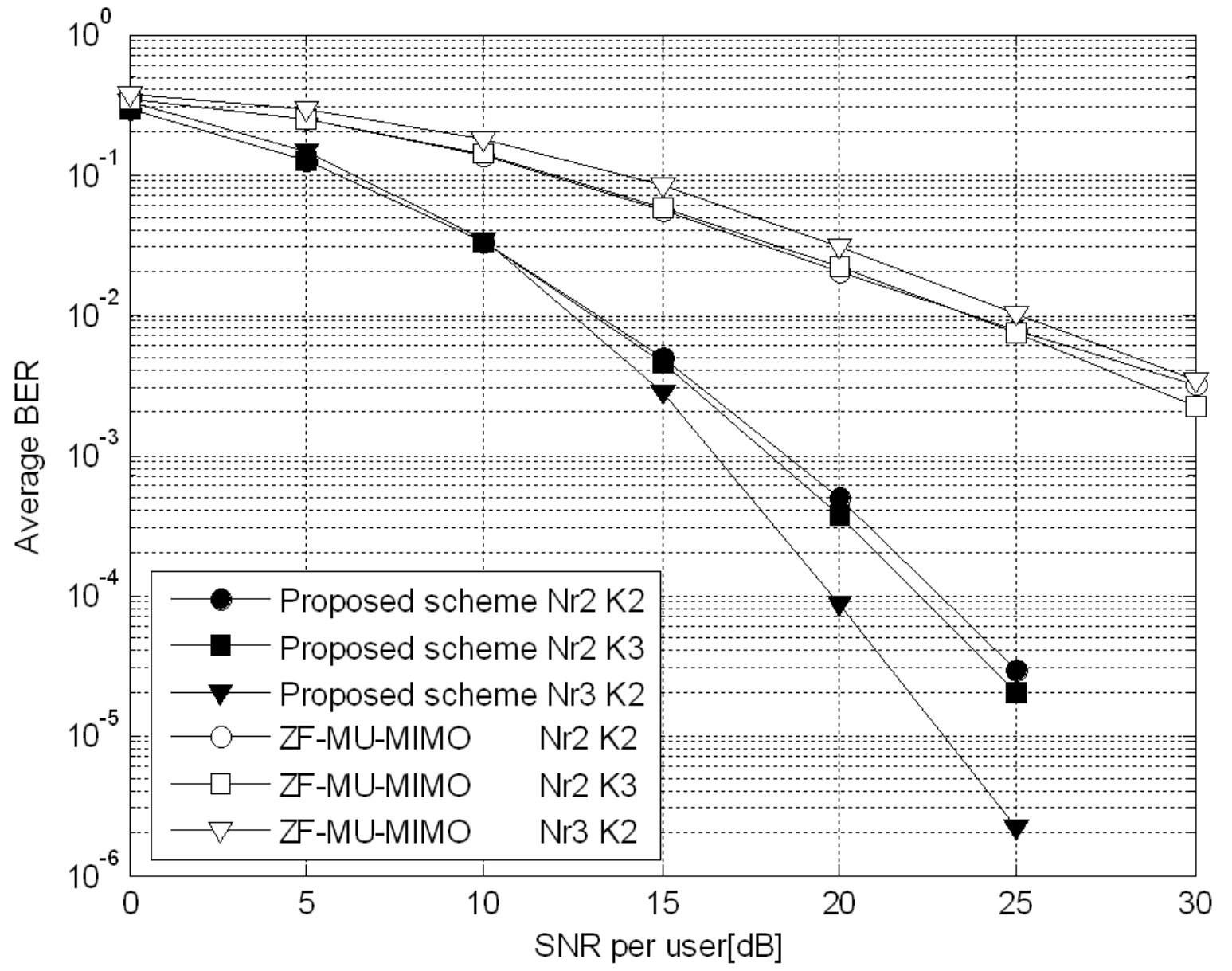

Fig. 7. BER performance comparison between the proposed scheme and the ZF precoding scheme without perturbing for 4-QAM.

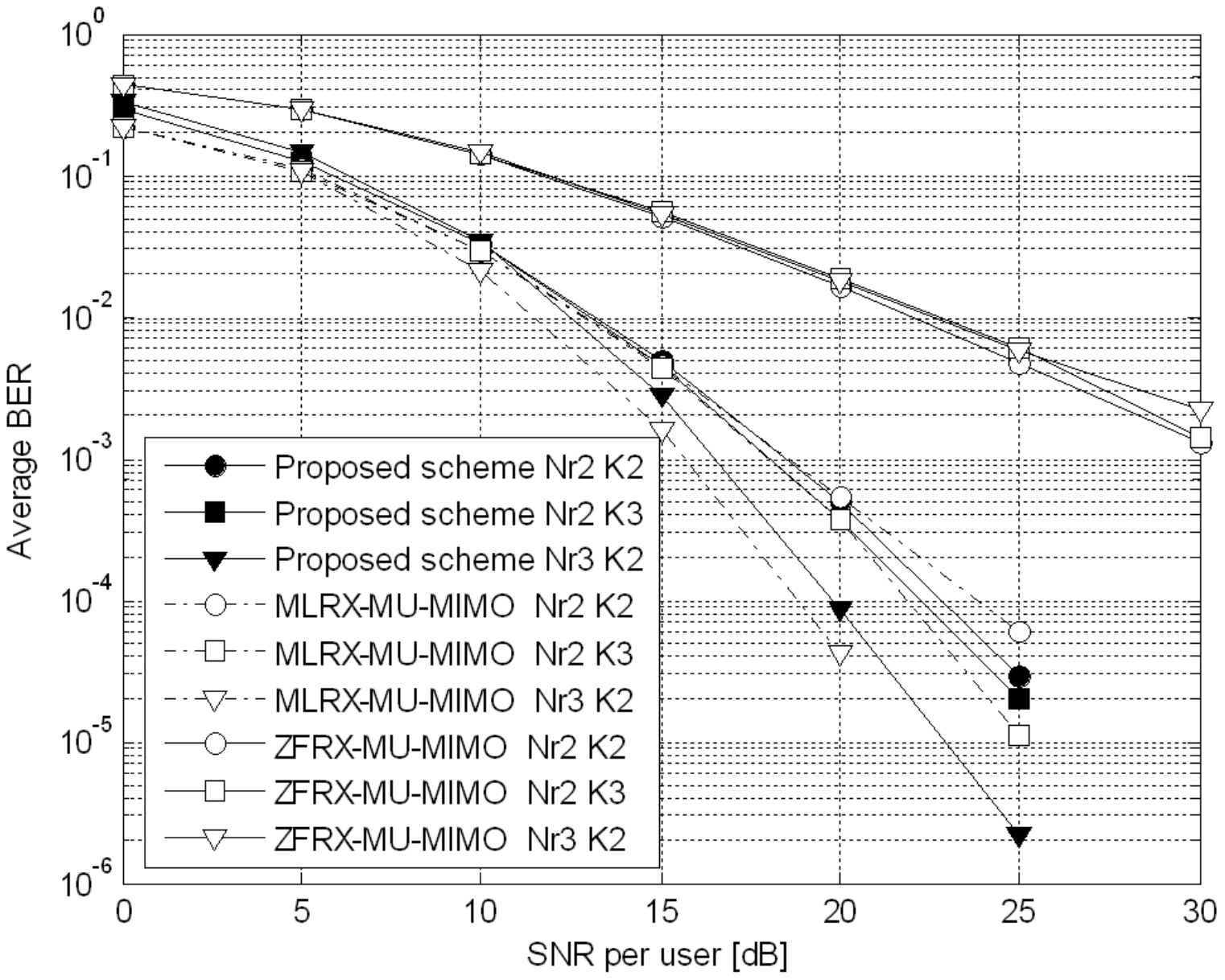

Fig. 8. BER performance comparison between the proposed scheme and the other schemes ([16]) which have perfect additional channel estimation at receiver for 4-QAM.

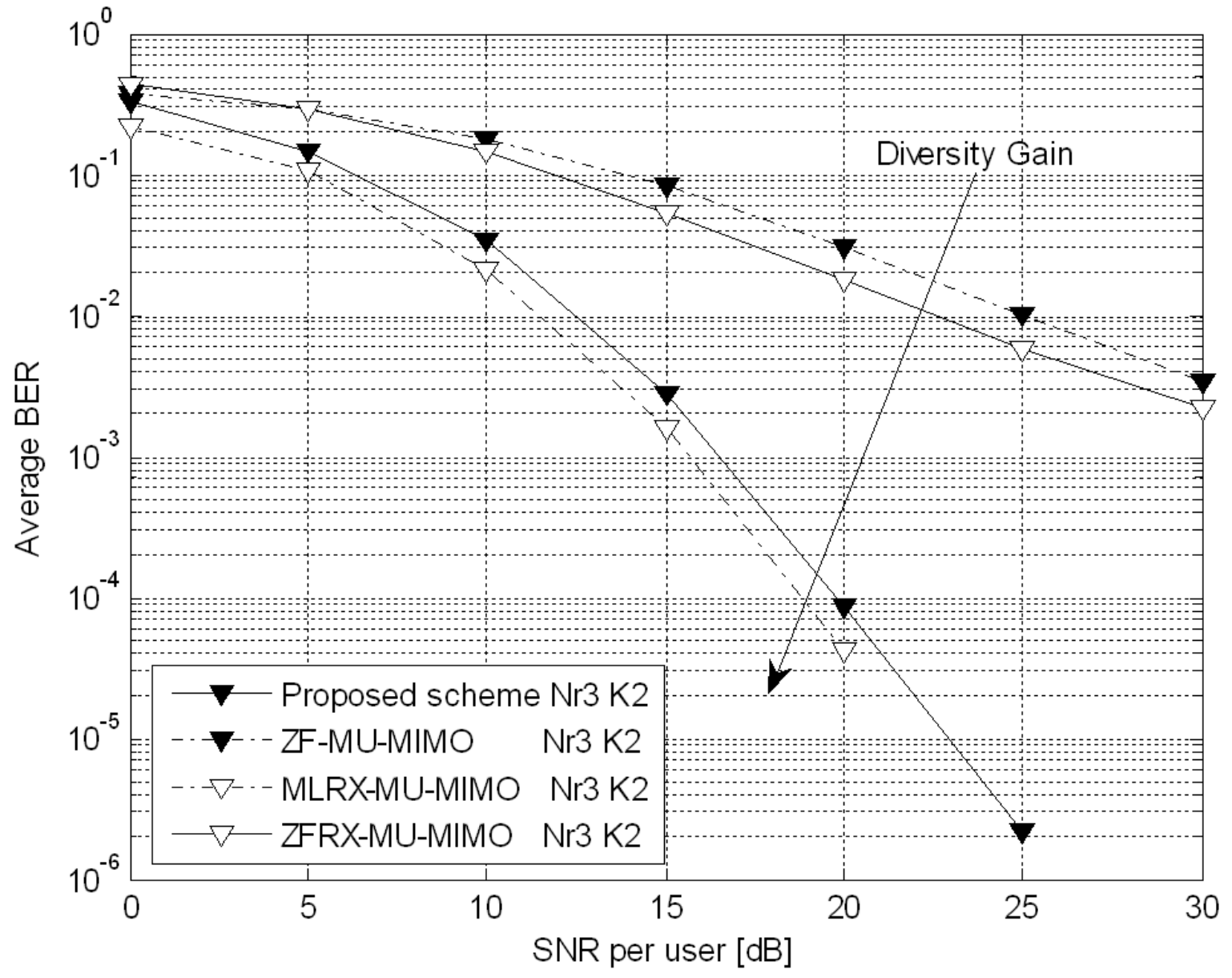

Fig. 9. Diversity gain comparison.